# Determining the quality evaluation procedures using the expert systems

Holban N., Ditoiu V., Iancu E.

**Abstract**— At this time, quality is a strategic instrument of the entities' global management, but it is also a determining element of their competitive spirit. The importance given to quality is abundantly found in the preoccupations of the European Union's Minister Board, by elaborating documents with a high impact over the quality of products/services in special, and organizations in general.

We live in an era, when the evolution of the social life puts the accent more and more on quality, resulted from various processes, at the level of various domains of the economical and social development.

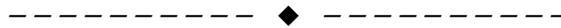

## 1 INTRODUCTION

As the increase and diversification of the products demand and industrial production development is more obvious, the notion of "product quality" continuously evolved, until present days when, under the condition of using electronic computers for quality control, this notion gained new meanings, such as "guided quality, assured quality, total quality, total quality management etc."

The non-stop improvement of quality is the current strategy type, able to approach in 2 ways[9]:

- *the classic way* – technological innovation strategy, involving major technological modifications, and as a result the product improvement is made in big jumps;
- *the modern way* – the Kaizen strategy; the term of Kaizen comes from Japanese, "kai" meaning change and "zen" meaning "for better, and is used under the meaning of "continuous improvement. According to the Kaizen concept, "If you take care of the quality, the profits will take care on their own!" But, prior to this is not result research, but the processes leading to these results and, implicitly, the workers. The results are obtained due to humans. The economical indices are consequences of the productive processes, and that is why the attention must be focused on the human resource.

The European Agreement Protocol concerning Conformity Evaluation and Industrial Products Acceptance, named PECA, is a protocol between the European Union and its member states, one on hand, and each of the candidate countries before their integration in the Union, on the other hand, for certain regulated domains, concerning mutual recognition of the conformity evaluation results and accepting the industrial products. The PECA is an instrument of the EU pre-integration strategy, and a support given to the candidate countries within the alignment process of the national legislations to the communitarian law.

The juridical basis of the PECA is assured by the stipulations of the European Agreement concerning co-operation in the industrial standards' domain and the possibility to make agreements of mutual recognition. The main clauses of the PECA refer to the producer's possibility to certify their products either in their home country, either in the European space, and the mutual acceptance of the conformity evaluation result and the legally made products, sold on one party's territory.

PECA aims for the products regulated by the communitarian law, based on the principles of the New European Approach, and its negotiation involves two main phases:

- making the frame protocol, containing the base principles of mutual recognition
- establishing sector annexes connected to the product groups which are to be introduced on the communitarian market and the candidate's, under the protocol- specified conditions.

The conditions to comply with in order to sign the PECA are:

- to develop and implement the horizontal law according to EU practices for standardization, metrology and infrastructure domains to evaluate conformity;
- to adopt and implement the specific technical legislation for products, in total compliance with the communitarian law by developing infrastructure to implement them, connected to adopting the harmonized European standards, evaluating product conformity and assuring legal stipulations respect control.

————————————————

- Holban N. *is from University of Suceava, 13 University Str., Suceava, Romania*
- Ditoiu V. *is from University of Suceava, 13 University Str., Suceava, Romania*
- Iancu E. *is from University of Suceava, 13 University Str., Suceava, Romania*

- To obtain the European Cooperation for Accreditation membership, and to make Multi-Lateral Agreements by the national accreditation organization;
- The national standardization organization has to be a member of the European Standardization Committee, and the European Committee for Electrotechnical Standardization (CENELEC);
- Temporarily close the negotiations afferent to the Free Merchandise Circulation Chapter.

As for the technical legislation for products, based on the principles of the New Approach in the domains making the object of PECA negotiations, it must be transposed in a total compliance with the communitarian law and implemented, or in an advanced implementation stage harmonized with the European practice.

According to the New Approach principles, the harmonized standards must offer a guaranteed protection level in fulfilling the essential requirements stipulated by the directives. In order to correctly apply these principles, conditions are necessary for a fair evaluation of compliance. For this, several guide principles were established [5]:

- applying conformity evaluation procedures and formulating criteria to use/choose these procedures;
- formulating criteria and procedures to evaluate, recognize/ designate and notify the product compliance evaluation organizations of the regulated domain;
- using the harmonized European standards concerning respect of the essential domains regulated by the New Approach directives;
- only the products complying with the essential requirements can be put on the market and set to function;
- using the harmonized European standards is voluntarily;
- using the harmonized European standards adopted as national standards offers the presumption of conformity with the matching essential requirements;
- the producers' possibility to choose between various compliance evaluation procedures stipulated by the applicable directive;
- promoting mutual recognition agreements concerning test and certification in the non-regulated domains;
- making mutual recognition agreements in order to stimulate commerce between the EU and the third-countries.

**The relation between the national and international standards**

This relation is one that involves certain specific functioning characteristics:
- adapting an international standard to a national one means publishing a national standard equivalent/ according to the international standard, or incorporating the international standard;
- applying an international standard for production, commerce, legislation etc. by a national standard, matching the subject of the international standard;
- the editing modification without changing the technical content of a standard compared to an international standard: correcting typos, using full stop instead of comma, adding an explication to avoid an eventual wrong interpretation of the original text; adding information or instructions;
- the reversibility principle- used to characterize situations in which what is acceptable in the international standard remains acceptable in the national one and vice versa;
- technical deviation of a national standard compared to an international standards concerning technical content.

There are two cases of technical deviations: - the major one, due to which the reversibility principle is not respected, that is- what is acceptable in the international standard becomes unacceptable in the national one and vice versa, - the major one, which does not affect the reversibility principle. For example, filling an attempt method standard with a normal form for the attempt declaration that is not included in the international standard.

At the project or product level, quality means:
- defining and obtaining the agreement on what must be reached and when (specifications and plans);
- defining the activities which must be performed and the afferent responsibilities (processes, procedures and standards)
- control, improvement and monitoring the activities and their results
- complying with the customer's requirements by properly applying the quality management system.

These targets are applicable in all the phases and development types of the IT products.

In making IT products, a great productivity is required, as in any other domain. This involves working in time limits, minimum cost and a product quality, which complies with the costumer's requirements. If these targets are not reached, we may consider that the product was made in an economic way. In order to obtain high productivity and comply with the requirements concerning budget, time limits and product quality, the general IT engineering principles must be respected, as well as the essential IT quality management's essential aspects.

The diagnostic is an investigation action carried out by a person or a group over a part of the organization, at the instance of competent authority [14]. Its purpose is to analyze the conformity of the functioning way and the management with the targeted objectives, to highlight incompliance and afferent causes, establishing a right action plan to meliorate the mentioned situation [13].

The base unit of the diagnostic is the instrument (an object made to interact with matter): diagnostic matrix, flow-chart, diagram, incompliance analysis graph etc. [14].

The intercession is an assembly of planned actions, used when the diagnostic occurs, in fact a directive idea applicable to the encountered situations [13]

The method is chosen according to the diagnostic's objectives and defines the instruments, content and stage succession, action principles. [9].

**The content and objectives of auditing**

According to SR EN ISO 19011, the audit is a systematic, independent and documented process, made to obtain audit proofs and their objective evaluation to determine how much the audit criteria are respected. The internal audits are lead/ for the organization itself, for the management analysis or other internal interests, and are the base for the conformity affidavit. In many cases, especially in the case of small organizations, independency can be demonstrated by the lack of responsibilities in the audited activity.

Quality audit can be made in order to: evaluate conformity of the process and the results of these processes (products, services) with a certain standard or another normative document; evaluate conformity of quality system elements or of the whole system with the specified rules; evaluate efficacy of the entity's quality system concerning reaching the fixed targets; identify the weak spots, sources of deficiencies in performing the activities in the enterprise; initiate corrective and improvement measures concerning the processes and results of these processes (products, services); monitoring the application of these established corrective and improvement measures.

One the other hand, the audits can be made to evaluate the quality system of an organization in relation to a certain standard or to verify if this system is implemented and complies with the pre-established requirements. By quality audit, the entity seeks also to identify the weak points, in order to reduce costs afferent to quality and monitoring the application of the established corrective measures.

Based on the auditors' results, the actions to improve the processes, products and management system's quality can be better established. As these audits are correctly planned and programmed, taking into account the nature and importance of the activities and are made by qualified auditors, the result of this entire audit process can be the continuous reduction of the deviations and the increase of the requirements compliance degree. Usually, the quality manual defines the quality system. The auditor in evaluating the quality system uses this document. The most efficient instrument the auditor can use is, in fact, the content of the quality manual.

The ISO 8402 standard defines quality audit as being *a systematic and independent examination, made to determine if the activity and their results, referring to quality, comply with the pre-established demands, if these demands are effectively implemented and corresponding to reaching the objectives.* At the moment, this definition is the most widely accepted.

Quality audits are [2], therefore, *"systematic"* examinations of the activities and their results referring to quality, being planned and programmed by the nature and importance of the mentioned activities. They are, on the other hand, *"independent"* examinations, that means they have to be made by persons, which are not involved in the audited domains.

By quality audit are evaluated:
- the entity's quality system as a whole, or elements of it;
- the entity's processes
- the results of the processes (products, services).

This evaluation is made in relation to the *"pre-established demands"* (applicable standards, quality manual, procedures, instructions, technical specifications etc.), to establish how much are they respected.

Quality audit does not resume only to establishing this correspondent, but seeks the demand's efficacy evaluation in reaching the objects assumed in the quality domain.

Based on the quality audit's results, the necessary *corrective actions* will be defined. These actions concern identifying and eliminating the found unconformities, in order to prevent them from repeating.

The corrective actions can involve procedure and quality system modifications, to assure quality improvement in each of the product's stages.

The audit must not be mistaken for *"quality supervision"* or *"inspection"*, which aim for keeping a process under control and accepting a certain product.

The efficacy of the quality auditors depends on the competence and experience of auditors. By the term of (quality) *auditor* is defined a person which has the necessary qualification to make quality audits. He must be authorized to make a certain audit.

The quality audit can be made in order to:
- evaluate conformity of the processes and results of these processes with a certain standard or another normative document;
- evaluate compliance of the quality system's elements, or the system as a whole with the specified requirements;
- evaluate efficacy of the entity's quality system concerning reaching for the targeted objectives;
- identify the weak spots, deficiency sources in performing activities in the entity;
- initiate corrective and improvement measures, concerning processes and results of these processes;
- monitor the application of the established corrective and improvement measures.

**The expert systems – quality evaluation instruments**

The IT component always connected to many other disciplines, the artificial intelligence, serves with finding the methods and means to build informational systems, which are to replicate the human capacities.

The Expert Systems (ES) can be part of a range of indispensable instruments used to make automatic or interactive systems, capable to make complex tasks [1]. The ES are part of a particular IT system class, based on artificial intelligence, aiming copy, with the help of com-puters, the knowledge and reasoning of the human experts.

The ES is a program that uses knowledge to obtain, as the human experts do, the results on activities difficult to examine. The main characteristic of the ES is derived from the knowledge base together with an algorithm specific to the reasoning method. An ES successfully solves problems for which a clear algorithmic solution does not exist or the algorithmic implementation is inefficient due its complexity.

Functionally, an ES is defined as a program that offers knowledge on obtaining results for the usual tasks difficult for the human experts to solve. Functionally and structurally, the ES have the following characteristics [8]:
- they are generally built to focus on the tasks with limited applicability range;
- there is an explicit separation between the reasoning knowledge and methods used to obtain knowledge- based conclusions;
- they are capable to explain own actions and judgment lines.

A main characteristic of the ES is the ones concerning their ability to feed with information on the reasoning used to get to the result. The multitude of the problems determines also the great volume of the knowledge base, bit an ES must be capable to solve in the same way the problems affected by uncertain and incomplete knowledge. In such situations, heuristic knowledge can be used, allowing finding the proper solution, but not necessarily the optimum one.

Although the ES term is often used in an equivalent way to the KBS one (knowledge base system), the latter has a wider including range. Therefore, for a KBS variations can appear in the knowledge base organization. Furthermore, a KBS can contain other subsystems. If the ES use the so-called *symbolic approach* of the AI, based on a symbolic processing of information, there also is the *connectionist approach,* of neuronal networks and the *evolutionist approach*, given by genetic algorithms, whereas KBS, the hybrid ones, there can be organized subsystems after such several approaches. These can create adequate interfaces for certain process types (for example, the neuronal networks are known to be used by the IA system in image processing or the genetic algorithms in certain optimizing problems classes).

*The knowledge base* is made of two components: the *rule base*, which includes general knowledge of the expertise domain, and the *fact base* made especially of specific knowledge for the problem that is to be solved.

In order to conceive and make an expert system, two personnel categories are required:
- the human experts, whose knowledge are to be collected and given to the users by the ES;
- the cogniticians, which assure transferring the human expert's reasoning strategies and knowledge in the specific structures of the method of representing knowledge and used informational instruments (logic programming, ES generator).
- With these are also included the traditional participants, the *final users*, which will exploit the system valuing its incorporated knowledge, *the designers and the programmers*, which will insure solving of the information problems involved by the make of ES.

Creating an ES prototype to determine the procedures to use in evaluating the quality level of a management system in relation to the requirements of the SR EN ISO 9001:2001 standard takes into account the expressed and implicit requirements of the standard and the applicability degree of this to the audited organization. For example, if we are to develop an expert system for quality testing, it will have to have information on the compliance level of certain criteria.

The minimal information we have to take into consideration when generating a base of knowledge are found in the quality manual, the documented procedures required in the SR EN ISO 9001:2001 standard, the organization's necessary documents to assure planning, operating and efficient control of its processes, demanded by the same standard.

In a brief analysis of the domain and the problem to solve, we have the following knowledge pieces:
1. Does the organization identify the processes necessary for the quality management system?
2. Does the organization determine the succession and interaction of these processes?
3. Does the organization determine the necessary criteria and methods to insure efficient operation and control of these processes?
4. Does the organization assure availability of necessary information and resources to operate and monitor these processes?
5. Does the organization monitor, measure and analyze these processes?
6. Does the documentation of the quality management system include: documented declarations of the politics concerning the quality domain and quality objectives?
7. Does the quality manual include the quality management system domain, including details and justifications for any exclusion?
8. Does the quality manual include documented procedures, established for the quality management system, or references to these?
9. Does the quality manual include a description of the interactions between the quality management system's processes?
10. Are the documents required by the quality management system controlled?
11. Is there a documented procedure established, to define the necessary controls for the approval, analysis and upgrade of the documents?
12. Is there established a documented procedure to define the necessary controls to identify the modifications and controlled distribution of the documents?
13. Are there established and maintained the recordings, to provide with proofs of compliance with the requirements and efficient functioning of the quality management system?

14. Is there a documented procedure established, to define the necessary control to identify and protect the keeping period of the recordings?

Based on this information, the rules may be built upon, which by going through the knowledge base will lead to answers used further on by the user.

**CONCLUSIONS**

The power lies in knowledge. Acquiring, formalizing and including the expert knowledge in the artificial system is the purpose of the Expert System's domain.

The expert system is a rather new informational technology, for which little experience exists. For this reason, we consider that a brief presentation of the most known existent design methods is not without interest, these methods applying different manners to conceive and realize.

The architecture of an expert system, far from rigid, is both modular, and flexible; it will be improved, as the utility of new modules will prove to be necessary.

The motivation of making a complex system for the quality's management and its testing, with the help of an expert system, is that it helps in:
- assisting the auditors during the evaluation of the management systems;
- assisting the managers making decisions in the quality domain, by a system thatows making scientifically- based decisions.

**REFERENCES**

[1] Andone, I., Tugui, A., (1999) – Sisteme inteligente în management, contabilitate, finante-banci si marketing, Editura Economica

[2]Ciurea, S.; Draganescu, N., *Managementul Calitii Totale*, Editura Economica, Bucureşti, 1995

[3]Constantinescu D., *Proiectarea şi implementarea sistemului caliatii*, in Tribuna calităţii, nr.1, 1998

[4]Dragulanescu N, *Politica europeana de promovare a calitatii*, in Qmedia, nr.1, 1999,

[5]Giaratano, J., Riley, G.- (2004*) Systems. Principles and Programming.Designing Expert Systems using CLIPS, Pub, Boston,*

[6] Holban, N. (coautor)– „*Modificări antropice ale mediului*", Editura Orizonturi Universitare,Timişoara, 2005; ISBN – 973-638-194-3

[7]Hinescu, A., *Cercurile de calitate ca instrumente ale managementului si al caliatii totale*, Tribuna Economica, nr. 26, 1996.

[8]Iancu E., Mates D., Voicu V., 2010 - **Considerations Regarding the Expert Systems in the Economy and the Use Method of the Production Systems Based on Rules,** Journal of Applied Computer Science & Mathematics, Ed. Universităţii, Suceava, ISSN 2066-4273

[9]Ionita I., *Managementul calitatii sistemelor tehnico-economice* , Ed. A.S.E., 2002

[10]Olaru M., *Managementul calităţii*, Editura Economică, Bucureşti, 1995

[10]Solomon G. şi A.Szuder, *Elemente de asigurarea calitatii*, Editura Tehnica, Bucuresti, 1998

[11] Stanciu, Ion , *Managementul calitatii totale (TQM)* Ed. Metropol, Bucuresti, 1996

[12]Ursachi I., *Management*, Atelier Poligrafic ASE, Bucureşti, 1993

[13]Colectia revistei Standardizarea, 2005-2009

[14]Colectia revistei Asigurarea Calitatii, 2000-2009

[15] http://www.iso9000.ro/rom/9001/iso9000guru.htm

[16] http://www.minind.ro/info_prod/Obiectiv1.html

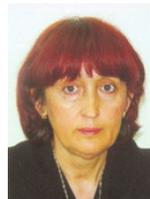

**Holban Nina**, **Professeur PhD** - Stefan cel Mare University of Suceava is a author of 3 specialty books, over 5 papers published in journals rated by ISI Thompson or within the international symposiums and conferences. She is a member in 3 international professional organizations. She's experienced in research contracts, she's part of the research team in 15 contracted projects, of which 11 are finalized and 4 are current

**Ditoiu Valeria- lecturer PhD -** Stefan cel Mare University, over 30 scientifically papers published in the country and abroad at the International Symposiums or Conferences. She is a member in 3 international professional organizations and scientifically referee in the editing committee, she's part of the research team in 5 contracted projects.

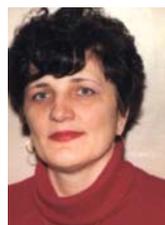

**Iancu Eugenia-** Stefan cel MareUniversity from Suceava. She is a doctorand at Technical Univesity from Timisoara. Authors 3 papers published in Journal rated by ISI Thompson and 25 scientifically papers published in the country and abroad at the International Symposiums or Conferences. She's experienced in research contracts, she's part of the research team in 9 contracted projects, of which 4 are finalized and 5 are current.

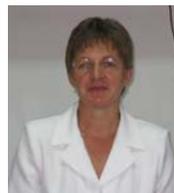